\documentclass[10pt]{article}

 \usepackage{amsmath}
\usepackage{graphicx}
\usepackage{epsfig}
\usepackage{color}
\usepackage{txfonts}
 \usepackage{amssymb}
 \usepackage{cite}

 \usepackage[labelfont=bf,labelsep=period,justification=raggedright]{caption}

 \makeatletter
\renewcommand{\@biblabel}[1]{\quad#1.}
\makeatother
 \date{}

\begin{document}

\begin{flushleft}
{\Large
\textbf{Sleep apnea-hypopnea quantification by cardiovascular data analysis}
}
\\
Sabrina Camargo$^{1,2,3\ast}$, 
 Maik Riedl$^1$,
 Celia Anteneodo$^{3,4}$,
 J\"urgen Kurths$^{1,5,6}$,
 Thomas Penzel$^{7}$, 
 Niels Wessel$^1$
 \\
\bf{1} Department of Physics, Humboldt-Universit\"at zu Berlin, Berlin, Germany \\
\bf{2} EMAp, Funda\c{c}\~ao Get\'ulio Vargas, Rio de Janeiro, Brazil \\
\bf{3} Department of Physics, PUC-Rio, Rio de Janeiro, Brazil  \\
\bf{4} National Institute of Science and Technology for Complex Systems, Rio de Janeiro, Brazil\\
\bf{5} Potsdam Institute for Climate Impact Research, Potsdam, Germany\\
\bf{6} Institute for Complex Systems and Mathematical Biology, University of Aberdeen, Aberdeen, United Kingdom\\
\bf{7} Sleep Center, Charit\'e University Hospital, Berlin, Germany \\
$\ast$ E-mail: sabrina.camargo@fgv.br

\end{flushleft}

\section*{Abstract}

Sleep apnea is the most common sleep disturbance and it is an important risk 
factor for cardiovascular disorders. Its detection relies on a polysomnography, 
a combination of diverse exams. 
In order to detect changes due to sleep disturbances such as sleep apnea occurrences, 
without the need of combined recordings, we mainly analyze systolic blood pressure signals  
(maximal blood pressure value of each beat to beat interval).
Nonstationarities in the data are uncovered by a segmentation procedure, 
which provides local quantities that are correlated to apnea-hypopnea events. 
Those quantities are the average length and average variance of stationary patches. 
By comparing them to an apnea score previously obtained by polysomnographic  
exams,  we propose an apnea quantifier based on blood pressure signal. 
This furnishes an alternative procedure for the detection of apnea based on a 
single time series, with an accuracy of $82\%$.

\section*{Introduction}

Sleep disturbances  (e.g., sleep apnea, insomnia, restless legs syndrome,
sleep walking, sleep terror) deserve serious attention  since they constitute
an important risk factor for cardiovascular disorders such as hypertension, cardiac
ischemia, sudden cardiac death, and stroke \cite{penzel.ea:07, stickgold:05,
smith:95}. 
Cardiorespiratory data, such as heart rate variability and respiratory
variability, can be useful to detect the  sleep disturbances, especially the
most common  sleep apnea.  Individuals suffering this kind of disorder usually
present daytime sleepiness, loud snoring and restless sleep.
A sleep apnea event is defined as a pause of the airflow lasting at least 10
secs.  If the air flow is lower than 50\% of normal, the resulting airflow
limitation is called a hypopnea \cite{penzel.ea:02}. 
When there is no inspiratory effort, then the event is classified as central.
If respiratory effort is made against an upper airway obstruction, then the
apnea event is classified as obstructive. Sleep apnea events can be also of a
mixed type.

In order to obtain a sleep profile the common practice is to combine records
collected by means of different exams: electroencephalography (EEG),
electromyography (EMG), and electro-oculography (EOG).  This set produces a
polysomnography, from which a visual scoring of sleep stages is evaluated,
assigning to each  stage  the pattern found in consecutive 30-second-long
epochs of the EEG, EMG, and EOG recordings. The resulting succession of
discrete sleep stages is referred to as hypnogram and supports diagnostic
decisions \cite{helland.ea:10}.

Signals of airflow, respiratory effort such as abdominal movement and oxygen
saturation of the blood are used in diagnosis of sleep apnea
\cite{caples.ea:05}, and as mentioned before, it requires combined records.
 Therefore, it would be desirable to evaluate
sleeping through an alternative procedure consisting of simpler data recordings.  
This is the goal we pursued in the present work. 

It is important to emphasize that cardiorespiratory time series 
are highly nonstationary, what restricts the use of standard tools. 
In that regard, Penzel et al. showed that changes in
heart rate variability in obstructive apneas were better quantified by
scaling analysis (using detrended fluctuation analysis) than by spectral
analysis \cite{al-hangari.ea:07, penzel.ea:03}. This is because, techniques
such as fast fourier transformation require stationarity in order to obtain a
meaningful estimation of the spectral components that are found in a time
series \cite{penzel4353742}. 
Hence, to deal with nonstationarity of  heart rate variability, as well as
of blood pressure variability, and consequently allowing the application of
techniques like spectral analysis, we apply a nonparametric segmentation
procedure to yield patches where stationarity is verified. In such locally
stationary segments, the statistical measures mean and variance
remain constant. Segmentation provides an indication of nonstationarity of
a time series, in particular, the intrinsic time scales. 
Moreover, by finding the stationary regimes, one might be able to identify
changes in time series, as those coming from the apnea occurrence.

\section*{Materials and Methods}

The study and the consent procedure were approved by the ethics
committee, Charit\'e - Universit\"atsmedizin, in Berlin. Participants provided their written
informed consent to participate in this study and the informed consent of all
subjects was recorded in paper form.
We analyzed data from 26 patients suffering from apnea-hypopnea, that is with
apnea-hypopnea index (AHI, the average number of apnea events per hour) larger
than 15, considering obstructive, central, and mixed sleep apnea events. 
The patients were then divided into two groups, according to their diurnal
systolic blood pressure levels: 10 hypertensive subjects (HT) and 16
normotensive patients (NT).  The mean values of the systolic blood pressure
were $142\pm4/93\pm8$ mmHg in HT and $120\pm10/81\pm7$ mmHg in NT. Moreover,
we add a control group  (C) with 7 non-apneic subjects.  All three groups are
age and sex matched, all male with mean age of $44.1\pm8.1$ years (HT),
$44.6\pm7.6$ (NT), and $44.8\pm6.7$ (C).  The apnea-hypopnea indexes in the
groups are $1.0\pm1.6$ (C), $42.5\pm24.0$ (NT) and $71.7\pm32.7$ (HT).

\begin{figure}[b!]
\begin{center}
\includegraphics[scale=1.0]{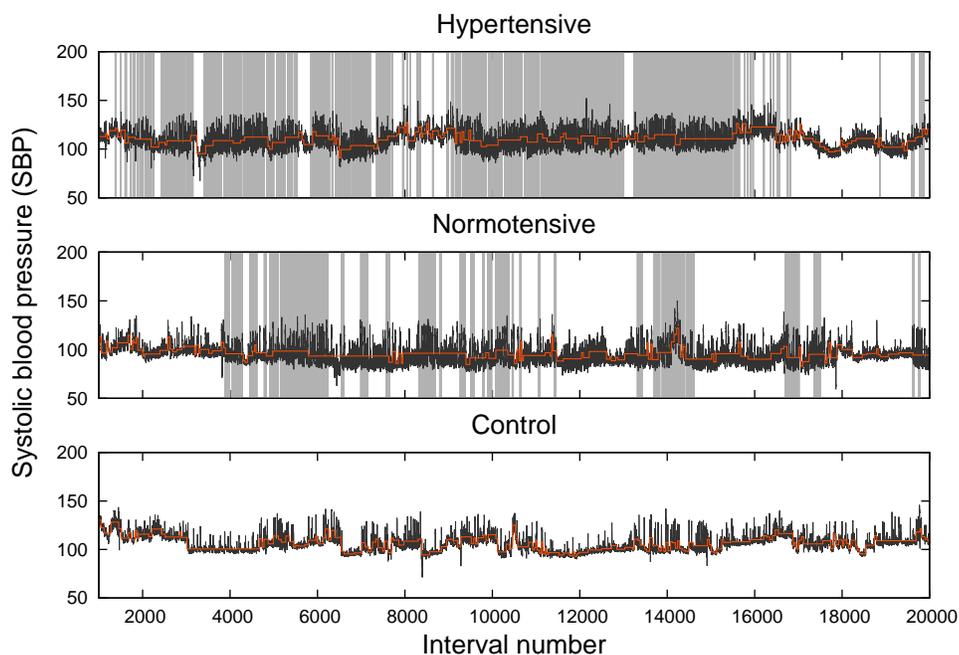}
\end{center}
\caption{Systolic blood pressure (SBP) time series (black line) 
and its local mean values from segmentation (light orange line) for  typical hypertensive 
(upper panel),  normotensive  (middle panel) and  control   (lower panel) subjects. 
Apnea events, detected via a polysomnography
examination, are represented by the light gray vertical lines. }
\label{fig:example_seg}
\end{figure}

The intervals between successive heartbeats (beat-to-beat intervals) were
extracted from  24h-electrocardiogram records \cite{suhrbier.ea:06}. To remove
artifacts caused by, e.g., premature beats, the beat-to-beat intervals were
filtered by means of interpolation using an adaptive filter
\cite{wessel.ea:07a}.  
Blood pressure was 'continuously' monitored (at a sampling rate of 200 Hz) 
with a  finger cuff sensor (Portapres Model2, BMI-TNO).
From the continuous blood pressure signal, 
the maximum value in each beat-to-beat interval was extracted, 
producing the time series of systolic blood pressure  (SBP) on a beat-to-beat
basis. Analogous procedure was followed by using minimum blood pressure values
to extract the beat-to-beat diastolic blood pressure (DBP) 
\cite{gapelyuk.ea:11,suhrbier.ea:10, penzel.ea:12}.
Beat-to-beat intervals   from blood pressure (BBI-BP)  records and from 
electrocardiograms (BBI-EKG) were also analyzed.  
Results for SBP and DBP series are similar, but SBP presents slightly better evaluation. 
We will concentrate on SBP in the further description.

We first dealt with the nonstationarity of the series performing a
segmentation into stationary-like patches based on the Kolmogorov-Smirnov (KS)
statistics, which provides a set of stationary data segments that compose
the signal \cite{camargo.ea:11}.  The KS-segmentation procedure is done
through the following steps: given the time series, all
points of the signal are considered as a potential cutting point, and we
compute the Kolmogorov-Smirnov distance
$D\equiv D_{KS}(1/n_{L}+1/n_{R})^{-1/2}$, 
between the cumulative distributions of the points belonging to the two
segments  placed at the left and the right sides of the cutting point, with
lengths $n_{L}$ and $n_{R}$, respectively. Thus, there will be one
value of $D$ corresponding to a hypothetical cut at each point of the signal,
and we determine the position $i_{max}$ where $D$ is maximal.
Once we know the position $i_{max}$ of the maximal distance $D$, $D^{max}$,
the statistical significance of this cut (at a desired significance level
$\alpha =1-P_{0}$) is verified by comparing $D^{max}$ with the result that
would be obtained by chance, given by the empirical curve 
$D^{max}_{crit}(n)=a(\ln n-b)^c$,
and $(a, b, c)$ = ($1.52$, $1.80$, $0.14$) for $P_0=0.95$, with $n=n_R+n_L$.
The signal is then split into two segments if $D^{max}$ exceeds its critical value
for the selected significance level $D^{max}_{crit}(n)$.  The procedure is
then applied to each one of the segments, starting from the full series
$\{x_{i},\;\;1\leq i\leq N\}$, where $N$ is the total number of data points,
until no segmentable patches are left. (See Refs.~\cite{camargo.ea:11,
camargo.ea:13b} for further details).
We performed the KS-segmentation with $\ell_0=30$, 
the minimal segment length, and $P_0=0.95$. The choice
of $\ell_0=30$ is based on its correspondence to the higher edge frequency of
the very low frequency   band of the heart rate with 0.03 Hz
\cite{task_force:96}.

Fig.~\ref{fig:example_seg} shows the time series of SBP (black lines) for
typical members  of  hypertensive (upper panel),  normotensive  (middle
panel), and control   (lower panel) groups,  with the first and second
patients suffering from sleep apnea-hypopnea.  The mean value of the data
segments provided by the KS-segmentation procedure is also represented (light
orange lines) to allow the reader to identify the stationary patches.  For
comparison, the sleep apnea events, detected via a polysomnography
examination, are represented by the light gray vertical lines.

\section*{Results}

\begin{figure}[h]
\begin{center}
\includegraphics[scale=1.0]{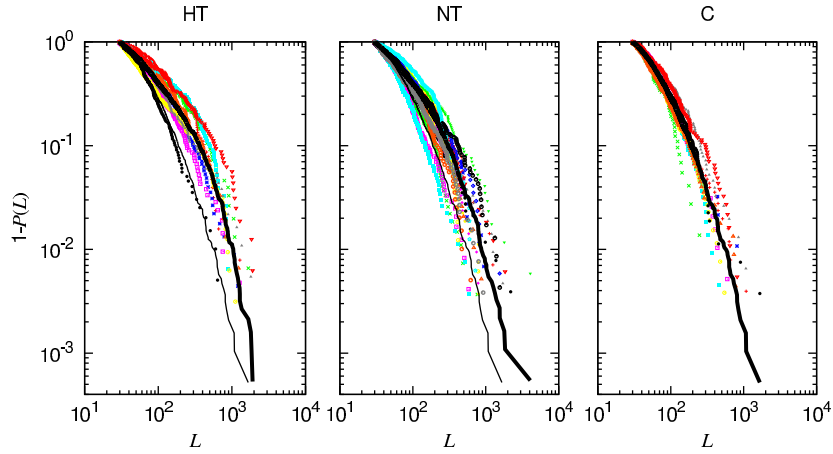}
\end{center}
\caption{Complementary cumulative distribution of segment lengths 
for each individual (color) and accumulated data of all
subjects in the same group (black solid line), for the hypertensive (HT), 
normotensive (NT) and control (C) groups. 
Drawn for comparison, the thin line in the first two panels (HT and NT) reproduces 
the control group accumulated data curve (C). }
\label{fig:cdf_l}
\end{figure}

Let us analyze the segmentation portraits of the SBP time series.  Concerning
the duration of stationary patches, we show in Fig.~\ref{fig:cdf_l}  the
complementary cumulative distribution of segment lengths. No substantial
differences are found between the distributions of the two apneic groups.  In
contrast, they differ from the control group distribution, where larger
patches are less frequent. 

\begin{figure}[t!]
\begin{center}
\includegraphics[scale=1.0]{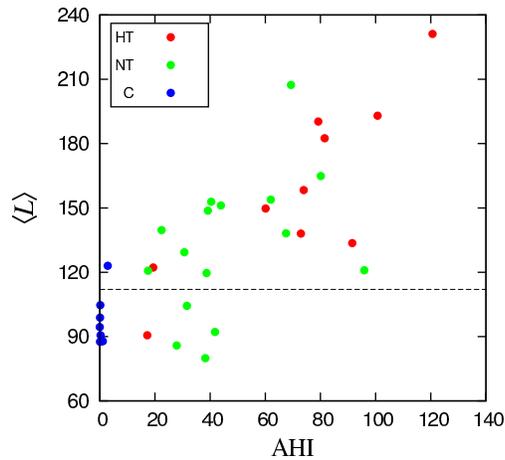}
\end{center}
\caption{Mean length of the segments $\langle L \rangle$ versus 
AHI for each subject. The dashed horizontal line represents the 
threshold value obtained by a ROC analysis. }
\label{fig:ahi_ell}
\end{figure}

In order to  inspect a possible correspondence between the typical 
duration of stationary segments and degree of apnea, 
we plot in Fig. \ref{fig:ahi_ell} the mean segment length $\langle L \rangle$ 
vs AHI, for each patient. As a matter of fact,  
a positive correlation between  $\langle L \rangle$ and AHI comes out (quantitatively, the 
Pearson coefficient is $r=0.77$). 
Moreover, one can set a threshold allowing to separate most apneic individuals. 
The threshold was chosen by minimizing the fractions of false negative and false positive 
results by means of ROC (receive operating characteristic) analysis
\cite{Metz1978}.
This threshold allows to identify above 80\% of apneic subjects. 
However, we will investigate if there are other quantities that might 
provide a better separation.

\begin{figure}[h]
\begin{center}
\includegraphics[scale=1.0]{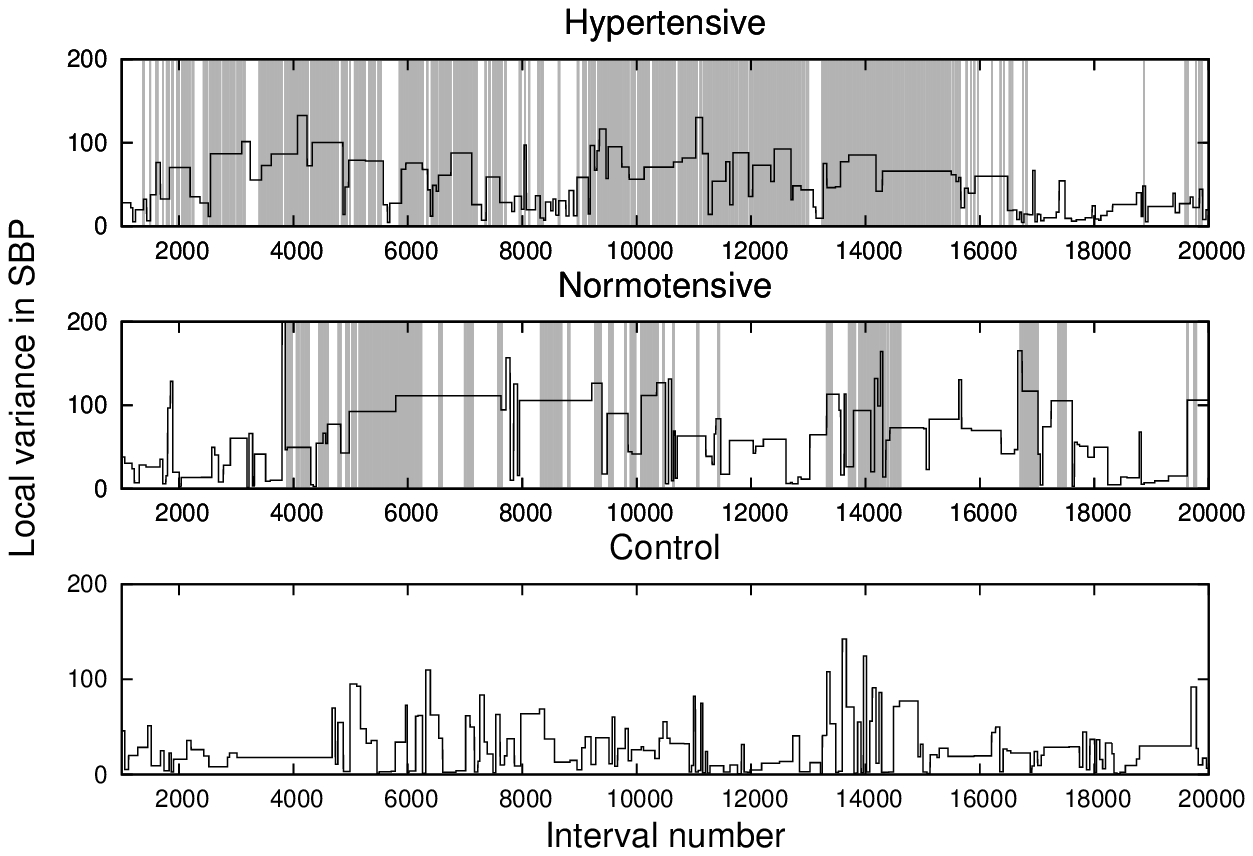}
\end{center}
\caption{Local variance (black lines) provided by the segmentation of SBP 
and the standard apnea detection represented by the light gray lines, for the same 
examples of Fig. \ref{fig:example_seg}.
}
\label{fig:example_seg_var}
\end{figure}

\begin{figure}[t!]
\begin{center}
\includegraphics[scale=1.0]{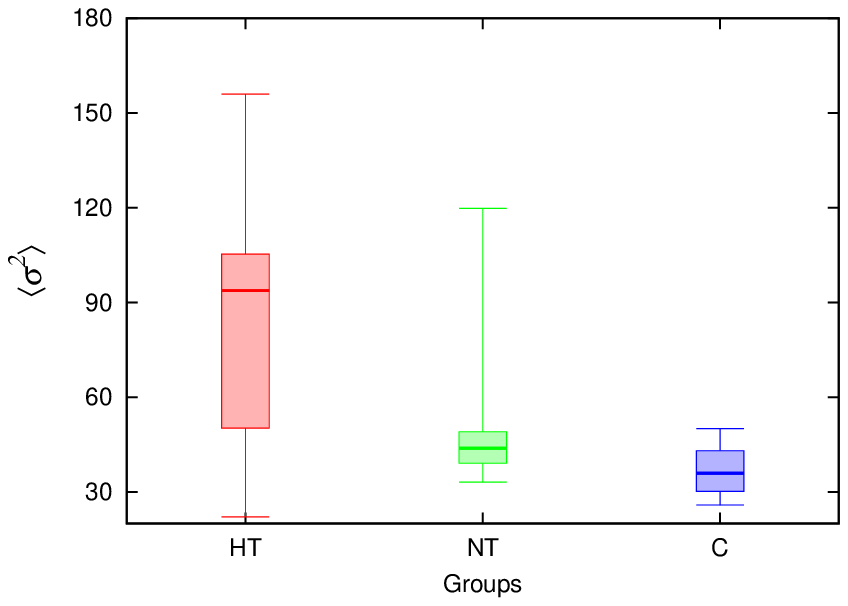}
\end{center}
\caption{Quartiles of the distribution of the average variance $\langle\sigma^2\rangle$, within each group. 
The horizontal lines limit the quartiles, the thicker one indicating the median.}
\label{fig:quartis}
\end{figure}

Let us now investigate  the statistics of stationary segments. 
 Since there is a tendency that apneic patients typically have 
larger mean variance (computed over all segments of each time series), 
then a natural  candidate for separation is, in principle, the variability in 
each segment. 
In Fig. \ref{fig:example_seg_var}, we depict the local variance $\sigma^2$ (variance of each patch)
for one individual of each group.
Notice that  apnea epochs have a strong influence on the
variability of the SBP signal, with increased dispersion (higher variance) in the gray areas, when
compared to the scored apnea, both for the hypertensive and normotensive cases. 

\begin{figure}[!h]
\begin{center}
\includegraphics[scale=1.0]{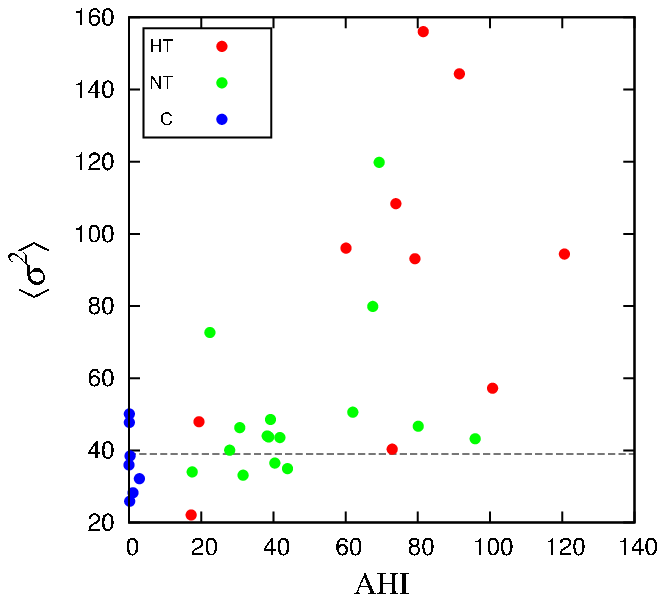}
\end{center}
\caption{Mean variance of the segments  $\langle\sigma^2\rangle$ versus  
AHI for each subject. The dashed horizontal line represents the ROC threshold.}
\label{fig:ahi_variance}
\end{figure}

Taking into account all the individuals of each group, 
we computed the quartiles (the three points that define the four equiprobable
intervals), for the average local variance $\langle \sigma^2 \rangle$, as shown in 
Fig.~\ref{fig:quartis}. 
Although there is a tendency that the control group presents lower dispersion, 
due to  overlapping, no clear group separation occurs by considering 
only the average variance for each patient. This is reinforced by the 
observation of a relatively weak correlation between the mean variance and AHI, 
as exhibited in Fig. \ref{fig:ahi_variance} (with $r=0.62$).

\begin{figure}[h!]
\begin{center}
\includegraphics[scale=1.0]{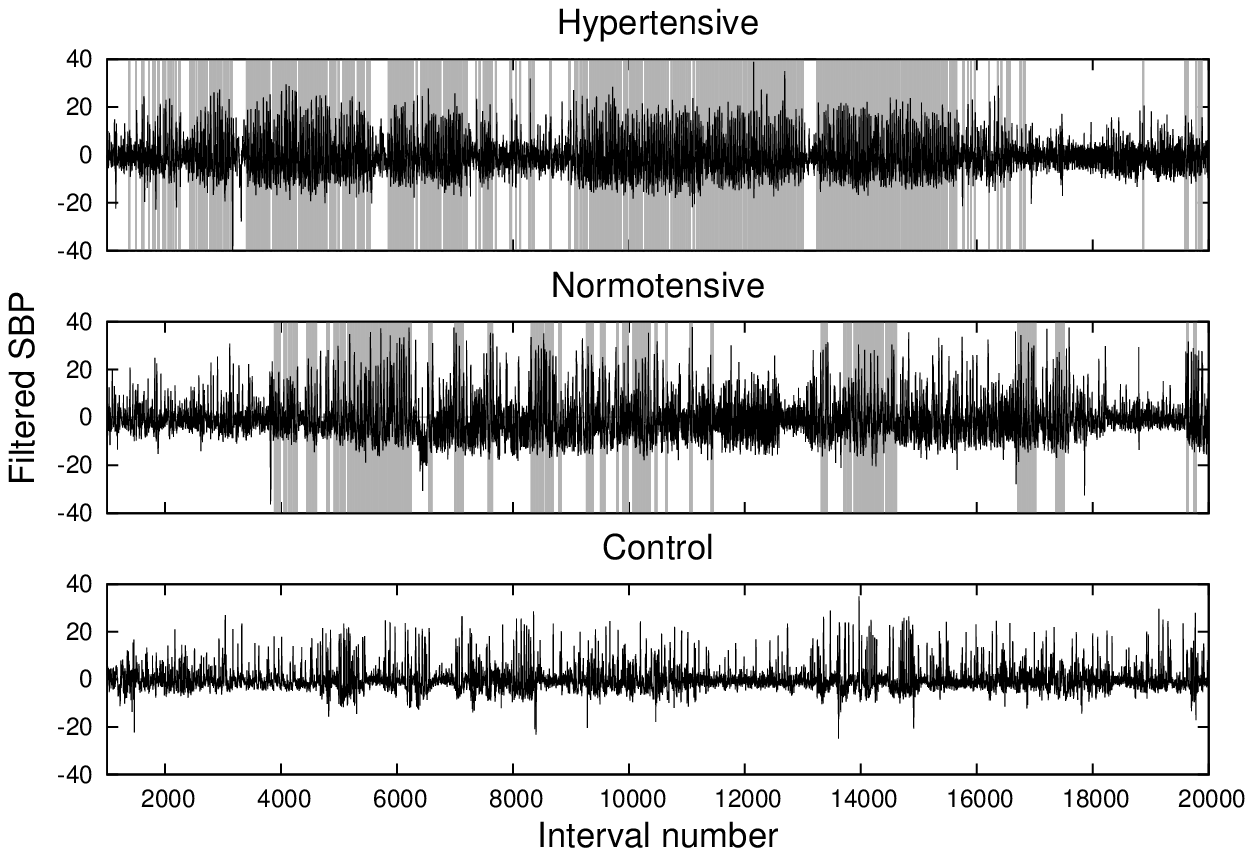}
\end{center}
\caption{Systolic blood pressure time series, once the local mean, $\mu$, was subtracted (black lines) and 
standard apneia detection (light gray vertical lines).}
\label{fig:diff_sbp}
\end{figure}
 
The average local mean $\langle \mu \rangle$ is  less efficient (with stronger overlap) 
than the average variance for separability purposes. The same arises when considering 
  higher order moments, for which  no significant differences amongst groups 
were detected.

The same analysis for the beat-to-beat series, BBI-BP and BBI-EKG, displayed weak 
correlations between the average length $\langle L \rangle$ and AHI, 
with Pearson coefficient $r=$ 0.11 and 0.05, respectively. 
Also a weak correlation   between the mean variance 
$\langle \sigma^2 \rangle$ and AHI was observed, with  $r=$  0.05 and 0.03, respectively. 
 Then,  in further analyses, we concentrated  on SBP, 
as a potential candidate for sleep apnea diagnosis.

The next step is to look at autocorrelations. In order to obtain a signal that
can be analyzed through standard spectral methods, we subtract the local mean
$\mu$ (the mean value of each data segment) from the corresponding patch,
yielding a filtered signal, as shown with black lines in
Fig.~\ref{fig:diff_sbp}.  
The removal of the local mean does not guarantee stationarity, as far as
variance (e.g., Fig \ref{fig:diff_sbp}) and higher-order moments may still
change. However, it furnishes a detrended signal more stationary than the
original one.

\begin{figure}[h!]
\begin{center}
\includegraphics[scale=1.0]{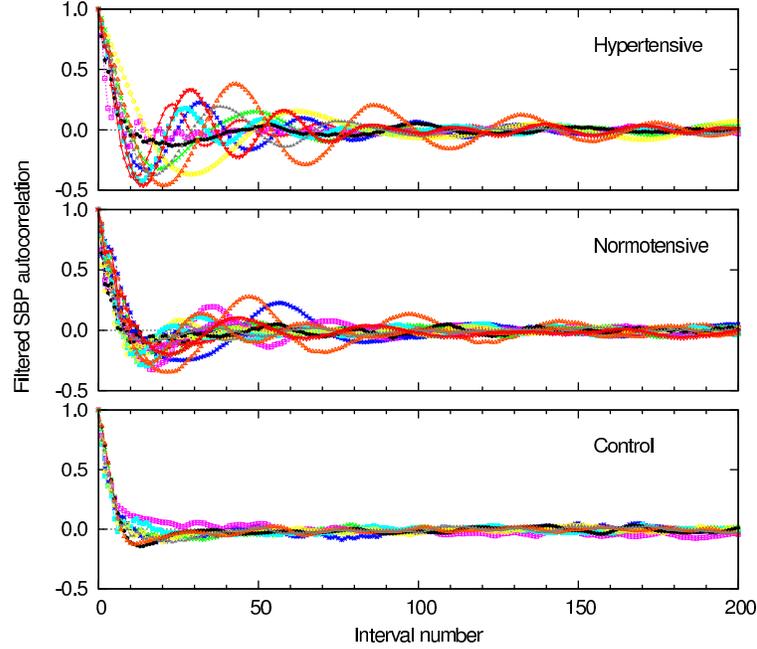}
\end{center}
\caption{Autocorrelation of the filtered SBP time series, computed for all the individuals 
of each group.}
\label{fig:autocor_diff}
\end{figure}

We then look at how the autocorrelation function of the filtered signal behaves for
each individual of each group, as shown in Fig.~\ref{fig:autocor_diff}. 
The autocorrelation function displays a noticeable behavior, with oscillatory patterns, 
which are more pronounced in subjects with sleep apnea, while rapidly
vanishing for the control group. Oscillations occur around zero, then alternating  
correlated and anticorrelated behavior.

\begin{figure}[t!]
\begin{center}
\includegraphics[scale=1.0]{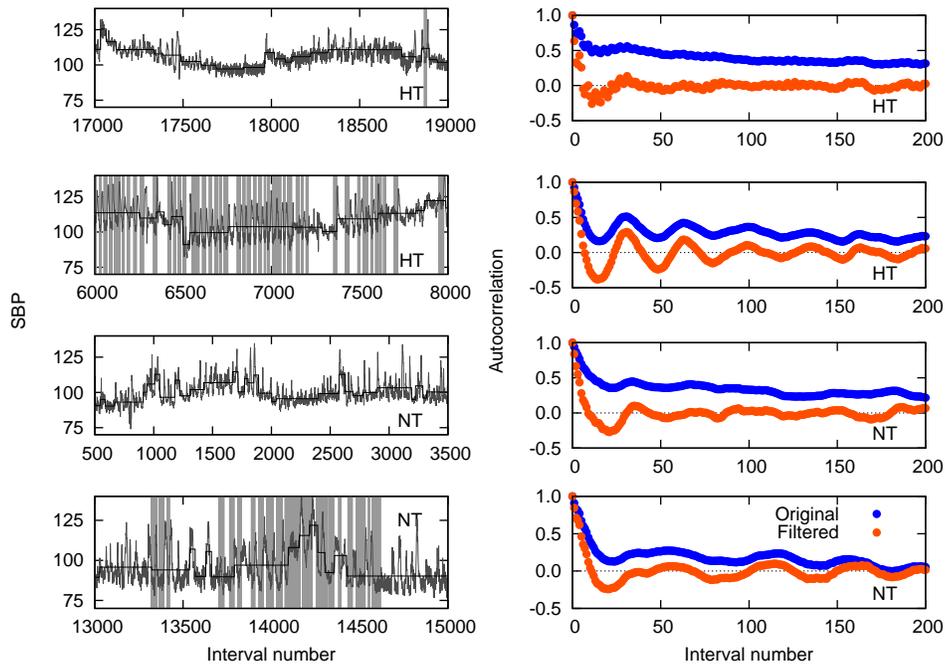}
\end{center}
\caption{Left panels: Patches during non-apnea   and
apnea epochs (recognizable by the absence/presence of gray vertical lines), for 
hypertensive and normotensive subjects. 
Right panels: the corresponding autocorrelation function for the original and 
filtered (local mean subtracted) signals.}
\label{fig:autocor_hyp}
\end{figure}

Let us remark that according to the apnea score, the hypertensive subjects are
in apnea during 28\%, on average, of the records, while the normotensive ones
are in apnea for 17\%, on average. In this way, in general, the effects of apnea may be
hidden, in particular in the case of correlations.
Then, we look at the autocorrelation function for two fragments of the SBP
time series: 2000 points during sleep apnea epochs and 2000 points in a
non-apnea epoch, in order to compare the effects of apnea in the same patient.
Fig.~\ref{fig:autocor_hyp} shows the  autocorrelation function for the
original and filtered (local mean subtracted) signals. We clearly find this
way that oscillations are related to apnea epochs. 
From the autocorrelation analysis, we conclude that 
the smaller amplitude of the oscillations in normotensive apneic subjects  
is not due to normal pressure but to a lower fraction of apnea epochs, 
then pointing to apnea as the source of the oscillations independently of the
blood pressure condition.

\begin{figure}[h!]
\begin{center}
\includegraphics[scale=1.0]{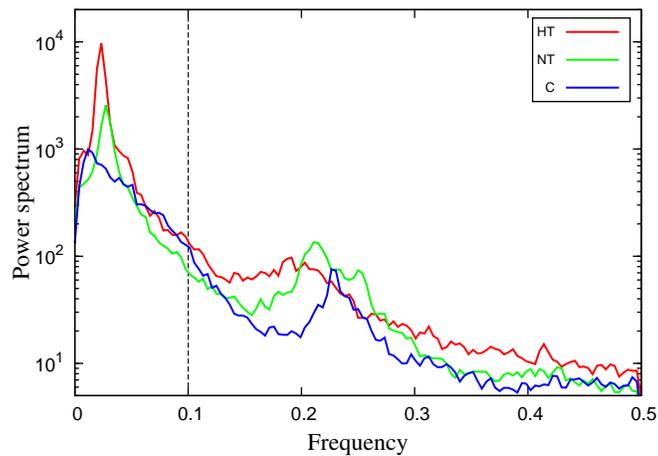}
\end{center}
\caption{Power spectrum of the filtered SBP signal, for a typical individual  of each group. 
Frequency corresponds to inverse interval number.}
\label{fig:spectral}
\end{figure}

\begin{figure}[t!]
\begin{center}
\includegraphics[scale=1.0]{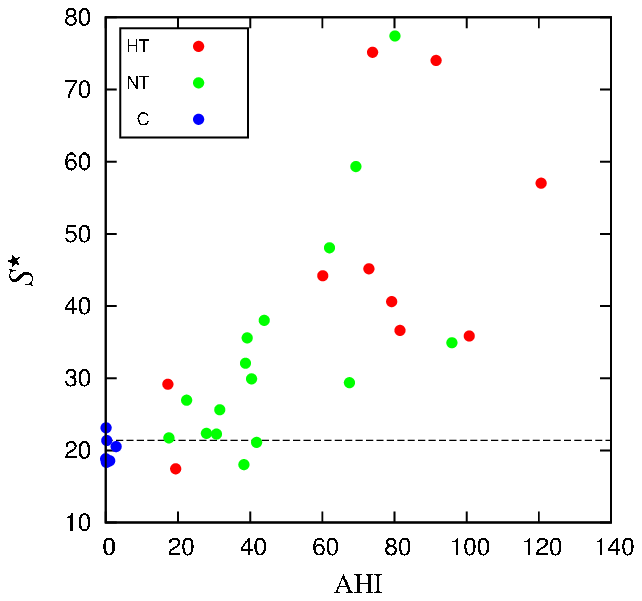}
\end{center}
\caption{Normalized maximum $S^\star$ of the power spectrum  versus AHI, for each individual. 
The dashed horizontal line represents the ROC threshold.}
\label{fig:max_area_1}
\end{figure}

In order to properly characterize the oscillations, we proceed to obtain the spectral 
density of the filtered signal, as illustrated in Fig. \ref{fig:spectral}. It manifests 
a main peak localized at a frequency about 0.02 in units of inverse interval number. 
In fact, the peak is typically more pronounced in apneic subjects. 
Then, according to the discussion in the precedent paragraph,  
concerning the lower amplitude of the oscillations in normotensive subjects, 
we define a relative amplitude  $S^\star$,  as 
the maximum value  normalized by the integral of the spectrum in the interval
$[0.0,0.1]$. 
In Fig. \ref{fig:max_area_1}, we represent the $S^\star$ against the apnea index, 
exhibiting the correlation between both quantities (with $r=0.74$). 

Spectral analysis of interbeat interval increments has been previously carried
out \cite{Roche1,Roche2}.  In that case, the relative percentage of the very
low frequency-component was taken as quantifier. However a ROC curve analysis
presented a worse classification than in our case. 

\begin{figure}[h!]
\begin{center}
\includegraphics[scale=1.0]{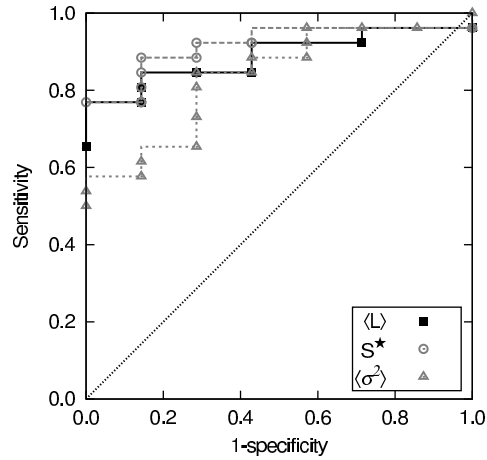}
\end{center}
\caption{ROC curves for $S^\star$, $\langle L \rangle$ and $\langle \sigma^2 \rangle$, obtained to identify 
patients with apnea. Threshold values (21.4, 112 and 39, respectively) shown in previous figures were obtained from the 
optimal classification corresponding to lowest distance to the upper left corner. 
}
\label{fig:ROC}
\end{figure}

The ROC curves for the three quantities here considered as potential
quantifiers, namely $(S^\star,\langle L \rangle,\langle \sigma^2 \rangle)$,
are displayed in Fig. \ref{fig:ROC}.  From those curves, the respective
thresholds were extracted.  The accuracy (the sum of true positive and true
negative subjects divided by the total sample size) of the optimal thresholds
$(S^\star,\langle L \rangle,\langle \sigma^2 \rangle)=(21.4, 112, 39)$  was
88, 82 and 82\%, respectively.

\begin{figure}[h!]
\begin{center}
\includegraphics[scale=1.0]{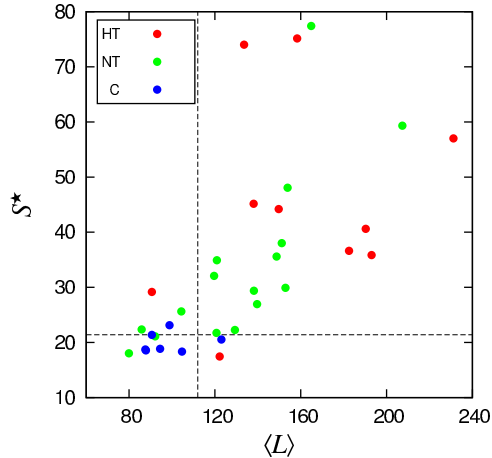}
\end{center}
\caption{Normalized maximum of the spectral density $S^\star$ vs 
mean segment length $\langle L \rangle$, for each individual. 
Dashed lines represent threshold values. }
\label{fig:diagram}
\end{figure}

The two quantities displaying higher Pearson coefficient, $r$, with respect to
the apnea index AHI are $S^\star$ and $\langle L \rangle$. Then we combine
them to obtain the diagram shown in Fig. \ref{fig:diagram}. 
We observe a neat separation of non-apneic subjects 
in the low $S^\star$ and low  $\langle L \rangle$ region.

%\newpage
\section*{Discussion}

We aim an alternative procedure for detecting apnea from simpler recordings than those 
composing a polysomnography. 
For that purpose, cardiovascular time series are good candidates.  
However, the nonstationarity of cardiovascular,
as well as cardiorespiratory, data hampers to employ standard spectral
techniques.  Therefore, we apply a segmentation procedure, that has been
recently developed \cite{camargo.ea:11}, to identify the patches of stationary behavior. 
Hence, for each patch, local quantities such as the statistical moments are
well defined. 
Moreover, the filtered signal, where the local mean was subtracted furnishes a
more stationary signal than the original one, hence it is more suitable, for
instance, for spectral analysis.

Through segmentation, we detected features of the series that are correlated
with apnea events.  These features are the relative intensity $S^\star$ of the
dominant  oscillations  in the autocorrelation function, the mean segment
length $\langle L \rangle$, and, in a less extent, the average variance
$\langle \sigma^2\rangle$. 
According to the ROC curves, each of these quantities already shows a better
performance than previous proposals based on heart rate
increments \cite{Roche1,Roche2}. Moreover, let us recall that obstructive
sleep apnea was found to be better quantified using blood pressure
variability, that are the data here considered, than heart rate increments
\cite{Roche1}.  The improvement may be attributed to our treatment of
nonstationarity, which is absent in the abovementioned studies.

Furthermore, the healthy subjects had less  larger patches than the patients
(Fig. \ref{fig:cdf_l}) which generally reflects a more active blood pressure
regulation in a normal person.  
Hence, the combination of mean segment length
and intensity of the oscillations allows to segregate apneic patients
efficiently, as shown in Fig. \ref{fig:diagram}.
This furnishes an alternative procedure for the detection of apnea from SBP
time series.

Let us also remark that a similar approach could be applied to other
physiological issues, where nonstationarity is a common place.


\begin{thebibliography}{10}

\bibitem{penzel.ea:07}
T.~Penzel, N.~Wessel, M.~Riedl, J.~W. Kantelhardt, S.~Rostig, M.~Glos,
  A.~Suhrbier, H.~Malberg, and I.~Fietze,
\newblock Cardiovascular and respiratory dynamics during normal and
  pathological sleep,
\newblock Chaos {\bf 17}(1), 015116 (2007).

\bibitem{stickgold:05}
R.~Stickgold,
\newblock Sleep-dependent memory consolidation,
\newblock Nature {\bf 437}, 1272 (2005).

\bibitem{smith:95}
C.~Smith,
\newblock Sleep states and memory processes,
\newblock Behav. Brain Res. {\bf 69}, 137 (1995).

\bibitem{penzel.ea:02}
T.~Penzel, J.~McNames, P.~de~Chazal, B.~Raymond, A.~Murray, and G.~Moody,
\newblock {Systematic comparison of different algorithms for apnoea detection
  based on electrocardiogram recordings},
\newblock {MEDICAL \& BIOLOGICAL ENGINEERING \& COMPUTING} {\bf {40}}({4}),
  {402--407} ({2002}).

\bibitem{helland.ea:10}
V.~C.~F. Helland, A.~Gapelyuk, A.~Suhrbier, M.~Riedl, T.~Penzel, J.~Kurths, and
  N.~Wessel,
\newblock Investigation of an automatic sleep stage classification by means of
  multiscorer hypnogram,
\newblock Methods Inf. Med. {\bf 49}, 467 (2010).

\bibitem{caples.ea:05}
S.~M. Caples, A.~S. Gami, and V.~K. Somers,
\newblock {Review Obstructive Sleep Apnea},
\newblock  {\bf 142}, 187--197 (2005).

\bibitem{al-hangari.ea:07}
H.~M. Al-Angari and A.~V. Sahakian,
\newblock Use of Sample Entropy Approach to Study Heart Rate Variability in
  Obstructive Sleep Apnea Syndrom,
\newblock IEEE TRANSACTIONS ON BIOMEDICAL ENGINEERING {\bf 54}, 1900 (2007).

\bibitem{penzel.ea:03}
T.~Penzel, J.~W. Kantelhardt, L.~Grote, J.-H. Peter, and A.~Bunde,
\newblock Comparison of Detrended Fluctuation Analysis and Spectral Analysis
  for Heart Rate Variability in Sleep and Sleep Apnea,
\newblock IEEE TRANSACTIONS ON BIOMEDICAL ENGINEERING {\bf 50}, 1143 (2003).

\bibitem{penzel4353742}
M.~Mendez, D.~Ruini, O.~Villantieri, M.~Matteucci, T.~Penzel, S.~Cerutti, and
  A.~Bianchi,
\newblock Detection of Sleep Apnea from surface ECG based on features extracted
  by an Autoregressive Model,
\newblock in {\em Engineering in Medicine and Biology Society, 2007. EMBS 2007.
  29th Annual International Conference of the IEEE}, pages 6105 --6108, 2007.

\bibitem{suhrbier.ea:06}
A.~Suhrbier, R.~Heringer, T.~Walther, H.~Malberg, and N.~Wessel,
\newblock Comparison of three methods for beat-to-beat-interval extraction from
  continuous blood pressure and electrocardiogram with respect to heart rate
  variability analysis,
\newblock Biomed. Tech. {\bf 51}, 70 (2006).

\bibitem{wessel.ea:07a}
N.~Wessel, H.~Malberg, R.~Bauernschmitt, and J.~Kurths,
\newblock Nonlinear methods of cardiovascular physics and their clinical
  applicability,
\newblock International Journal of Bifurcation and Chaos {\bf 17}, 3325 (2007).

\bibitem{gapelyuk.ea:11}
A.~Gapelyuk, M.~Riedl, A.~Suhrbier, J.~Krämer, G.~Bretthauer, H.~Malberg,
  J.~Kurths, T.~Penzel, and N.~Wessel,
\newblock Cardiovascular regulation in different sleep stages in the
  obstructive sleep apnea syndrome,
\newblock Biomed Tech (Berl) {\bf 56}, 207 (2011).

\bibitem{suhrbier.ea:10}
A.~Suhrbier, M.~Riedl, H.~Malberg, T.~Penzel, G.~Bretthauer, J.~Kurths, and
  N.~Wessel,
\newblock Cardiovascular regulation during sleep quantified by symbolic
  coupling traces,
\newblock Chaos {\bf 20}, 045124 (2010).

\bibitem{penzel.ea:12}
T.~Penzel, M.~Riedl, A.~Gapelyuk, A.~Suhrbier, G.~Bretthauer, H.~Malberg,
  C.~Sch\"obel, I.~Fietze, J.~Heitmann, J.~Kurths, and N.~Wessel,
\newblock Effect of CPAP therapy on daytime cardiovascular regulations in
  patients with obstructive sleep apnea,
\newblock Comput Biol Med. {\bf 42}, 328 (2012).

\bibitem{camargo.ea:11}
S.~Camargo, S.~M.~D. Queir\'os, and C.~Anteneodo,
\newblock Nonparametric segmentation of nonstationary time series,
\newblock Phys. Rev. E {\bf 84}, 046702 (2011).

\bibitem{camargo.ea:13b}
S.~Camargo, M.~Riedl, C.~Anteneodo, N.~Wessel, and J.~Kurths,
\newblock Diminished heart beat nonstationarities in congestive heart failure,
\newblock Frontiers in Physiology {\bf 7}, 107 (2013).

\bibitem{task_force:96}
T.~Force,
\newblock Guidelines - Heart rate variability,
\newblock European Heart Journal {\bf 17}, 354 (1996).

\bibitem{Metz1978}
C.~E. Metz,
\newblock {Basic Principles of ROC Analysis},
\newblock Seminars in Nuclear Medicine {\bf VIII} (1978).

\bibitem{Roche1}
L.~Poupard, I.~Court-Fortune, V.~Pichot, F.~Chouchou, J.-C. Barthélémy, and
  F.~Roche,
\newblock Use of high-frequency peak in spectral analysis of heart rate
  increment to improve screening of obstructive sleep apnoea,
\newblock Sleep and Breathing {\bf 15}, 837--843 (2011).

\bibitem{Roche2}
F.~Roche, E.~Sforza, D.~Duverney, J.-R. Borderies, V.~Pichot, O.~Bigaignon,
  G.~Ascher, and J.-C. Barthélémy,
\newblock Heart rate increment: an electrocardiological approach for the early
  detection of obstructive sleep apnoea/hypopnoea syndrome,
\newblock Clin. Sci. {\bf 107}, 105--110 (2004).

\end{thebibliography}
\end{document}